\documentclass[preprint,showpacs,superscriptaddress,amsmath,amssymb,showkeys]{revtex4}

\usepackage{graphicx}% Include figure files
\usepackage{dcolumn}% Align table columns on decimal point
\usepackage{bm}% bold math

\begin{document}

%\setlength{\textheight}{7.7truein}  %for 2nd page o
%\def\bea{\begin{eqnarray}}
%\def\eea{\end{eqnarray}}
%\def\nn{\nonumber}
%\draft
%\preprint{}

\title{Four-neutrino analysis of 1.5km-baseline reactor antineutrino oscillations}

%\thispagestyle{empty}
%\begin{titlepage}
\author{Sin Kyu Kang} \email{skkang@seoultech.ac.kr}
\affiliation{School of Liberal Arts, Seoul National University
of Science and Technology, Seoul 139-743, Korea}
\affiliation{Pittsburgh Particle Physics, Astrophysics, and Cosmology Center,
Department of Physics and Astronomy, University of Pittsburgh, Pittsburgh, PA 15260, USA}
\author{Yeong-Duk Kim} \email{ydkim@sejong.ac.kr}
\affiliation{Department of Physics, Sejong University, Seoul, 143-747, Korea}
\author{Young-Ju Ko} \email{yjko4u@gmail.com}
\author{Kim Siyeon} \email{siyeon@cau.ac.kr}
\affiliation{Department of Physics, Chung-Ang University, Seoul, 156-756, Korea}

\date{August 14, 2014}
\begin{abstract}

The masses of sterile neutrinos are not yet known, and depending on the orders of magnitudes, their existence may explain reactor anomalies or the spectral shape of reactor neutrino events at 1.5km-baseline detector. Here, we present four-neutrino analysis of the results announced by RENO and Daya Bay, which performed the definitive measurements of $\theta_{13}$ based on the disappearance of reactor antineutrinos at km-order baselines. Our results using 3+1 scheme include the exclusion curve of $\Delta m^2_{41}$ vs. $\theta_{14}$ and the adjustment of $\theta_{13}$ due to correlation with $\theta_{14}$. The value of $\theta_{13}$ obtained by RENO and Daya Bay with a three-neutrino oscillation analysis is included in the $1\sigma$ interval of $\theta_{13}$ allowed by our four-neutrino analysis.
\end{abstract}

%\end{titlepage}
\maketitle \thispagestyle{empty}

%%%%%%%%%%%%%%%%%%%%%%%%%%%%%%%%%%%%%%%%%%%%%%%%%%%%%%%%%%%%

\section{Introduction}

Understanding of the Pontecorvo-Maki-Nakagawa-Sakata(PMNS) matrix\cite{PMNS} is now moving to another stage, due to the determination of the last angle by multi-detector observation of reactor neutrinos at Daya Bay\cite{An:2012eh} and RENO\cite{Ahn:2012nd}, whose success was strongly expected from a series of oscillation experiments, (T2K\cite{Abe:2011sj}, MINOS\cite{Adamson:2011qu}, Double Chooz\cite{Abe:2011fz,Abe:2012tg}), which all contributed to the forefront of neutrino physics \cite{Beringer:1900zz}. A number of 3$\nu$ global analyses\cite{Fogli:2012ua,Tortola:2012te} have presented the best fit and the allowed ranges of masses and mixing parameters at 90\% confidence level(CL) by crediting RENO and Daya Bay for the definitive measurements of $\sin^22\theta_{13}$. For instance, the best-fit values given in the analysis of Fogli et al.\cite{Fogli:2012ua} are $\Delta m_{21}^2= 7.5\times10^{-5}\mathrm{eV}^2, ~ \sin^2\theta_{12}=3.2\times10^{-1}, ~\Delta m_{32}^2=2.4\times10^{-3}\mathrm{eV}^2, ~\sin^2\theta_{13}=2.8\times 10^{-2},$ and $\sin^2\theta_{23}=4.8\times 10^{-1}$ for normal hierarchy. While all three mixing angles are now known to be different from zero, the values of the CP violating phases are completely unknown. Although there are a number of global analysis which presented consistent values of masses and mixing parameters \cite{Fogli:2012ua,Tortola:2012te,GonzalezGarcia:2012sz}, we focus on $\theta_{13}$ and its associated factors obtained by RENO and Daya Bay.

Although the three-neutrino framework is well established phenomenologically, we do not rule out the existence of new kinds of neutrinos, which are inactive so-called sterile neutrinos. Over the past several years, the anomalies observed in LSND \cite{Aguilar:2001ty}, MiniBooNE \cite{AguilarArevalo:2007it}, Gallium solar neutrino experiments \cite{gallium} and some reactor experiments\cite{Declais:1994su} have been partly reconciled by the oscillations between active and sterile neutrinos. In a previous work, we also examined whether the oscillation between sterile neutrinos and active neutrinos is plausible, especially when analyzing the first results released from Daya Bay and RENO \cite{Kang:2013gpa}. There are also other works with similar motivations \cite{Bora:2012pi}.

After realizing the impact of the large size of $\theta_{13}$, both reactor neutrino experiments have continued and updated the far-to-near ratios and $\sin^22\theta_{13}$. Daya Bay improved their measurements and explained the details of the analysis. RENO announced an update with an extension until October 2012, and modified their results as follows: The ratio of the observed to the expected number of neutrino events at the far detector $R=0.929$ replaced the former value of $R=0.920$, and $\sin^22\theta_{13}=0.100$ replaced the former best fit of $\sin^22\theta_{13}=0.113$\cite{seo:nutel}. The spectral shape was also modified. Again, we examine the oscillation between a sterile neutrino and active neutrinos in order to determine whether four-neutrino oscillations are preferred to three-neutrino oscillations. This work is focused on $\Delta m_{14}^2$ within the range of $\mathcal{O}(0.001\mathrm{eV}^2)$ to $\mathcal{O}(0.1\mathrm{eV}^2)$, where $\Delta m_{14}^2$ oscillations might have appeared in the superposition with $\Delta m_{13}^2$ oscillations at far detectors of $\mathcal{O}(1.5\mathrm{km})$ baselines. Since the mass of the fourth neutrino is unknown, it is worth verifying its existence at all available orders of magnitude which are accessible from  different baseline sizes. For instance, the near detector at RENO can search reactor antineutrino anomalies with $\Delta m_{14}^2\sim\mathcal{O}(1\mathrm{eV}^2)$\cite{Mention:2011rk, Mueller:2011nm}.

This article is organized as follows: In Section II, the survival probability of electron antineutrinos is presented in four-neutrino oscillation scheme. We exhibit the dependence of the oscillating aspects on the order of $\Delta m_{41}^2$, when reactor neutrinos in the energy range of 1.8 to 8 MeV are detected after travel along a km-order baseline. In Section III, the curves of the four-neutrino oscillations are compared with the spectral shape of data through October 2012 to search for any clues of sterile neutrinos and to see the changes in $\sin^22\theta_{13}$ due to the coexistence with sterile neutrinos. Broad ranges of $\Delta m_{41}^2$ and $\sin^22\theta_{14}$ remain. In the conclusion, the exclusion bounds of $\sin^22\theta_{14}$ and the best fit of $\sin^22\theta_{13}$ are summarized, and the consistency between rate-only analysis and shape analysis is discussed.

\section{Four neutrino analysis of event rates in multi detectors}

The four-neutrino extension of unitary transformations from mass basis to flavor basis is given in terms of six angles and three Dirac phases:
    \begin{eqnarray}
    \widetilde{U}_\mathrm{F} &=& R_{34}(\theta_{34})R_{24}(\theta_{24},\delta_2)R_{14}(\theta_{14}) \cdot \nonumber \\
    & & \cdot R_{23}(\theta_{23})R_{13}(\theta_{13},\delta_1)R_{12}(\theta_{12},\delta_3),
    \label{4by4trans}
    \end{eqnarray}
where $R_{ij}(\theta_{ij})$ denotes the rotation of the $ij$ block by an angle of $\theta_{ij}$.
When a 3+1 model is assumed as the minimal extension, the 4-by-4 $\widetilde{U}_\mathrm{F}$ is given by
\begin{widetext}
    \begin{eqnarray}
    \widetilde{U}_\mathrm{F} &=&
	\left(\begin{matrix}	
        c_{14} & 0 & 0 & s_{14} \\
		-s_{14}s_{24} & c_{24} & 0 & c_{14}s_{24} \\
		-c_{24}s_{14}s_{34} & -s_{24}s_{34} & c_{34} & c_{14}c_{24}s_{34} \\
		-c_{24}c_{34}s_{14} & -s_{24}c_{34} & -s_{34} & c_{14}c_{24}c_{34}
	    \end{matrix}\right)
    \left( \begin{matrix}
    U_{e1} & U_{e2} & U_{e3} & 0 \\
    U_{\mu1} & U_{\mu2} & U_{\mu3} & 0 \\
    U_{\tau1} & U_{\tau2} & U_{\tau3} & 0 \\
    0 & 0 & 0 & 1
        \end{matrix} \right)
         \\
    &=& \left( \begin{matrix}
    c_{14}U_{e1} & c_{14}U_{e2} & c_{14}U_{e3} & s_{14} \\
    \cdots & \cdots & \cdots & c_{14}s_{24} \\
    \cdots & \cdots & \cdots & c_{14}c_{24}s_{34} \\
    \cdots & \cdots & \cdots & c_{14}c_{24}c_{34}
        \end{matrix} \right),
\end{eqnarray}
\end{widetext}
where the PMNS type of a 3-by-3 matrix $U_\mathrm{PMNS}$ with three rows, $(U_{e1} ~ U_{e2} ~ U_{e3}), ~ (U_{\mu1} ~ U_{\mu2} ~ U_{\mu3})$ and $(U_{\tau1} ~ U_{\tau2} ~ U_{\tau3}),$ is imbedded. The CP phases $\delta_2$ and $\delta_3$ introduced in Eq.(\ref{4by4trans}) are omitted for simplicity, since they do not affect the electron antineutrino survival probability at the reactor neutrino oscillation.

The survival probability of $\bar{\nu}_e$ produced from reactors is
    \begin{eqnarray}
    P_\mathrm{Th}(\bar{\nu}_e\rightarrow\bar{\nu}_e) &=& |\sum_{j=1}^4|\widetilde{U}_{ei}|^2 \exp{i\frac{\Delta m_{j1}^2L}{2E_\nu}}|^2 \\
        &=& 1- \sum_{i<j}4|\widetilde{U}_{ei}|^2|\widetilde{U}_{ej}|^2\sin^2(\frac{\Delta m_{ij}^2L}{4E_\nu}),
    \end{eqnarray}
where $\Delta m_{ij}^2$ denotes the mass-squared difference $(m_i^2-m_j^2)$. It can be expressed in terms of combined $\Delta m_{ij}^2$-driven oscillations as
    \begin{eqnarray}
    P_\mathrm{Th}(\bar{\nu}_e\rightarrow\bar{\nu}_e) &=& 1-c_{14}^4c_{13}^4\sin^2 2\theta_{12}\sin^2(1.27\Delta m_{21}^2\frac{L}{E}) \nonumber \\
    &-& c_{14}^4\sin^2 2\theta_{13}\sin^2(1.27\Delta m_{31}^2\frac{L}{E})  \label{pee} \\
    &-& c_{13}^2 \sin^2 2\theta_{14}\sin^2(1.27\Delta m_{41}^2\frac{L}{E}) \nonumber \\
    &-& s_{13}^2 \sin^2 2\theta_{14}\sin^2(1.27\Delta m_{43}^2\frac{L}{E}), \nonumber
    \end{eqnarray}
where $\Delta m_{32}^2 \approx \Delta m_{31}^2$ and $\Delta m_{42}^2 \approx \Delta m_{41}^2$.  The size of $m_4$ relative to $m_3$ is not yet constrained.
The above $P_\mathrm{Th}$ is understood only within a theoretical framework, since the energy of the detected neutrinos is not unique but is continuously distributed over a certain range. So, the observed quantity is established with a distribution of neutrino energy spectrum and an energy-dependent cross section. Analyses of neutrino oscillation averages accessible energies of the neutrinos emerging from the reactors. The measured probability of survival is
    \begin{eqnarray}
    \langle P \rangle = \frac{\int P_{\mathrm{Th}}(E) \sigma_{\mathrm{tot}}(E) \phi(E) dE}{\int  \sigma_{\mathrm{tot}}(E) \phi(E) dE}, \label{avg_prob}
    \end{eqnarray}
where $\sigma_{\mathrm{tot}}(E)$ is the total cross-section of inverse beta decay(IBD), and $\phi(E)$ is the neutrino flux distribution from the reactor. The total cross section of IBD is given as
    \begin{eqnarray}
    \sigma_{\mathrm{tot}}(E)=0.0952 \left( \frac{E_e\sqrt{E_e^2-m_e^2}}{1\mathrm{MeV}^2}\right)\times 10^{-42}\mathrm{cm}^2,
    \end{eqnarray}
where $E_e \approx E_\nu-(M_n-M_p)$\cite{Mueller:2011nm, Vogel:1999zy}. The flux distribution $\phi(E)$ from the four isotopes $(\mathrm{U}^{235},\mathrm{Pu}^{239},\mathrm{U}^{238},\mathrm{Pu}^{241})$ at the reactors  is expressed by the following exponential of a fifth order polynomials of $E_\nu$
    \begin{eqnarray}
    \phi(E_\nu)=\exp\left(\sum_{i=0}^5 f_i E_\nu^i \right), \label{flux}
    \end{eqnarray}
where $f_0=+4.57491 \times 10, f_1=-1.73774 \times 10^{-1}, f_2=-9.10302 \times 10^{-2}, f_3=-1.67220 \times 10^{-5}, f_4=+1.72704 \times 10^{-5}$, and $f_5=-1.01048 \times 10^{-7}$ are obtained by fitting the total flux of the four isotopes with the fission ratio expected at the middle of the reactor burn up period \cite{Huber:2011wv}.

\begin{figure}
\resizebox{80mm}{!}{\includegraphics[width=0.75\textwidth]{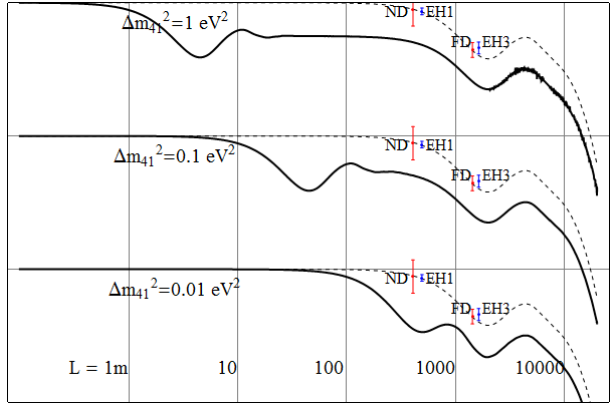}}
\caption{\label{fig1:distance} The dependency of $\langle P \rangle$ in Eq.(\ref{avg_prob}) on $\Delta m_{41}^2$ is presented, when $\Delta m_{31}^2=2.32\times 10^{-3}\mathrm{eV}^2$ as taken in RENO and Daya Bay. Typical shapes of $\langle P \rangle$ vs. distance are drawn for comparison with the measured ratios in the two experiments. The amplitudes of $\Delta m_{31}^2$ oscillation and $\Delta m_{41}^2$ oscillation are given by $\sin^22\theta_{13}=0.10$ and $\sin^22\theta_{14}=0.10$, respectively, as an example. }
\end{figure}

\begin{figure}
\resizebox{70mm}{!}{\includegraphics[width=0.75\textwidth]{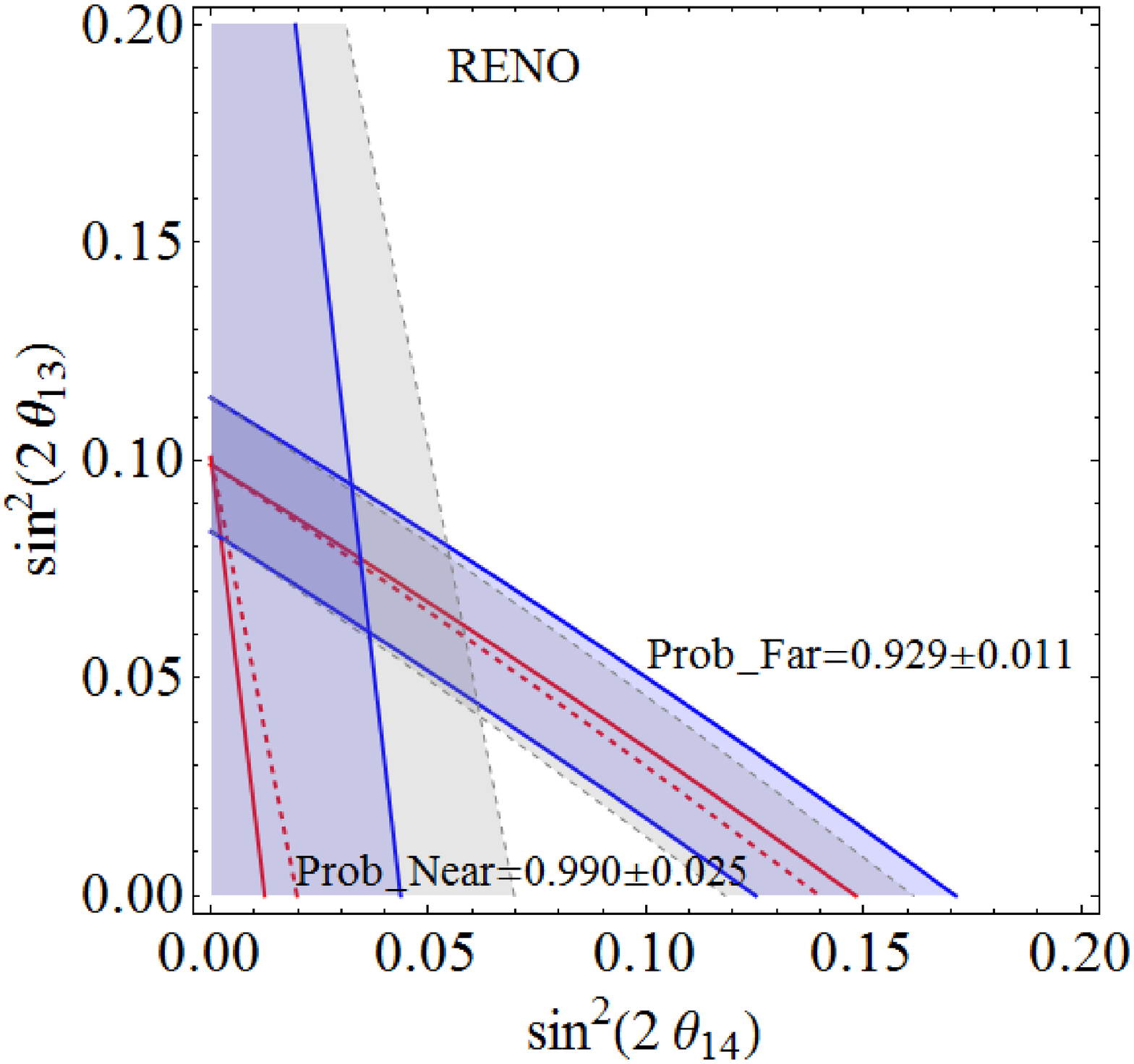}}
\resizebox{70mm}{!}{\includegraphics[width=0.75\textwidth]{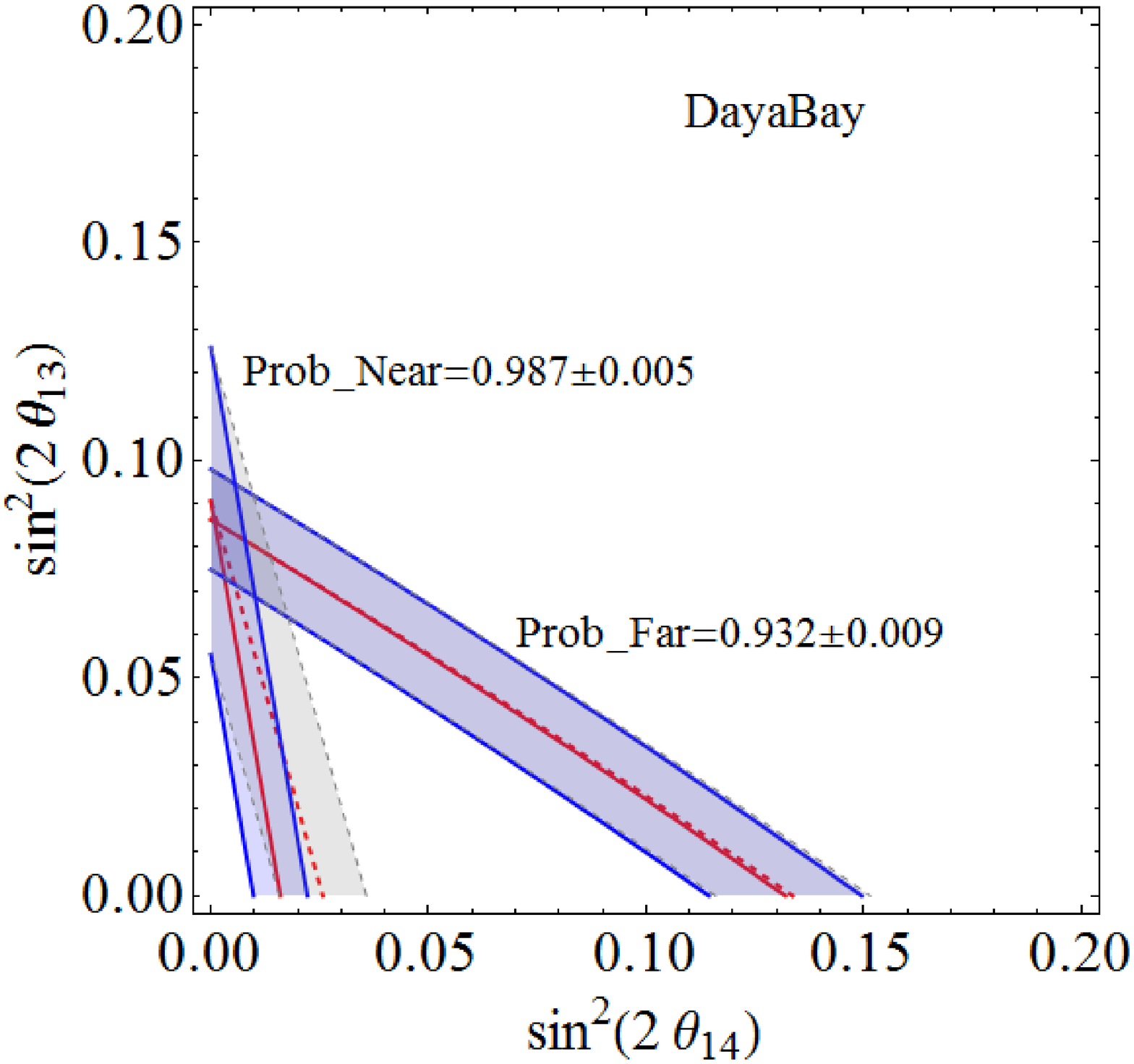}}
\caption{\label{fig:theta13theta14} Four-neutrino analysis for the observed to expected ratios at both ND and FD of RENO. The shaded areas indicate the available combination of $\sin^22\theta_{13}$ and $\sin^22\theta_{14}$ which is compatible with the ranges of $R_\mathrm{Near}$ and $R_\mathrm{Far}$ released by RENO and Daya Bay. Blue(gray) areas are obtained with $\Delta m_{41}^2=0.01(0.1)\mathrm{eV}^2$. The $4\nu$ analysis of the far-to-near ratio at Daya Bay is also given for comparison.}
\end{figure}

The curves in Fig.\ref{fig1:distance} show $\langle P \rangle$ as $L$ increases in a logarithmic manner, where the three patterns of probabilities are shown according to the order of $\Delta m_{41}^2$. The first bump in each curve corresponds to the oscillation due to $\Delta m_{41}^2$, while the second bump that appears near $1500 \mathrm{m}$ corresponds to the oscillation due to $\Delta m_{31}^2$. RENO and Daya Bay were designed to observe the $\Delta m_{31}^2$-driven oscillations at far detector(FD) according to three-neutrino analysis, while additional detector(s) at a closer baseline perform the detection of neutrinos in the same condition. The comparison of the number of neutrino events at FD to the number of events at the near detector(ND) is an effective strategy to determine the disappearance of antineutrinos from reactors. That is, the $\sin^22\theta_{13}$ is evaluated by the slope of the curve between ND and FD, while their absolute values of event numbers do not affect the estimation of the angle $\theta_{13}$. Both experiments used the normalization to adjust the data to satisfy the boundary condition which is that there is no oscillation effect before the ND. From Fig. \ref{fig1:distance}, it can be shown that the magnitude of $\Delta m_{41}^2$ can affect not only the normalization factor but also the ratio between the FD and ND.

The flux-weighted average baselines of the near detector(ND) and the far detector(FD) of RENO, $\overline{L}_\mathrm{near}$ and $\overline{L}_\mathrm{far}$, are 407.3m and 1443m, respectively. The baselines of Daya Bay, named EH1, EH2, and EH3, have lengths of EH1=494m, EH2=554m, and EH3=1628m, respectively, so that, conventionally, EH1 and EH2 are regarded as near detectors while EH3 is regarded as a far detector.

After the first release of results, Daya Bay and RENO updated the far-to-near ratio of neutrino events with additional data. Daya Bay reported a ratio of $R=0.944\pm0.007(\mathrm{stat})\pm0.003(\mathrm{syst})$ with $R(\mathrm{EH1})=0.987\pm0.004(\mathrm{stat})\pm0.003(\mathrm{syst})$\cite{An:2012bu}. RENO also reported an update with additional data from March to October in 2012, where $R(\mathrm{FD})=0.929\pm0.006(\mathrm{stat})\pm0.009(\mathrm{syst})$\cite{seo:nutel}. Their measurements are marked in Fig.\ref{fig1:distance}. In three-neutrino analysis, the far-to-near ratios give the $\Delta m^2_{31}$-oscillation amplitude $\sin^22\theta_{13}=0.089\pm0.010(\mathrm{stat})\pm0.005(\mathrm{syst})$ and $\sin^22\theta_{13}=0.100\pm0.010(\mathrm{stat})\pm0.015(\mathrm{syst})$ in Daya Bay and RENO, respectively. On the other hand, the far-to-near ratio and the measured-to-expected ratio are understood as a combination of $\Delta m^2_{31}$ oscillations and $\Delta m^2_{41}$ oscillations as shown in Fig.\ref{fig1:distance}. For a given value of $\Delta m^2_{41}$, $0.01\mathrm{eV}^2$ or $0.01\mathrm{eV}^2$, the combination of $\sin^22\theta_{14}$ and $\sin^22\theta_{13}$ is described in Fig.\ref{fig:theta13theta14}. In the case of RENO, the $\langle P (\Delta m^2_{41}=0.01\mathrm{eV}^2) \rangle$ and $\langle P (\Delta m^2_{41}=0.1\mathrm{eV}^2) \rangle$ curves which pass the error bars at ND and FD are drawn as blue(gray)-shaded areas. The area where the two shaded areas, ND and FD, overlap is the allowed region in $\sin^22\theta_{13}-\sin^22\theta_{14}$ space using rate-only analysis. The corresponding analysis for Daya Bay is shown together in Fig.\ref{fig:theta13theta14}. The value of $\sin^22\theta_{13}$ is in good agreement with the results released by the two experiments.

\begin{figure}
\resizebox{80mm}{!}{\includegraphics[width=0.75\textwidth]{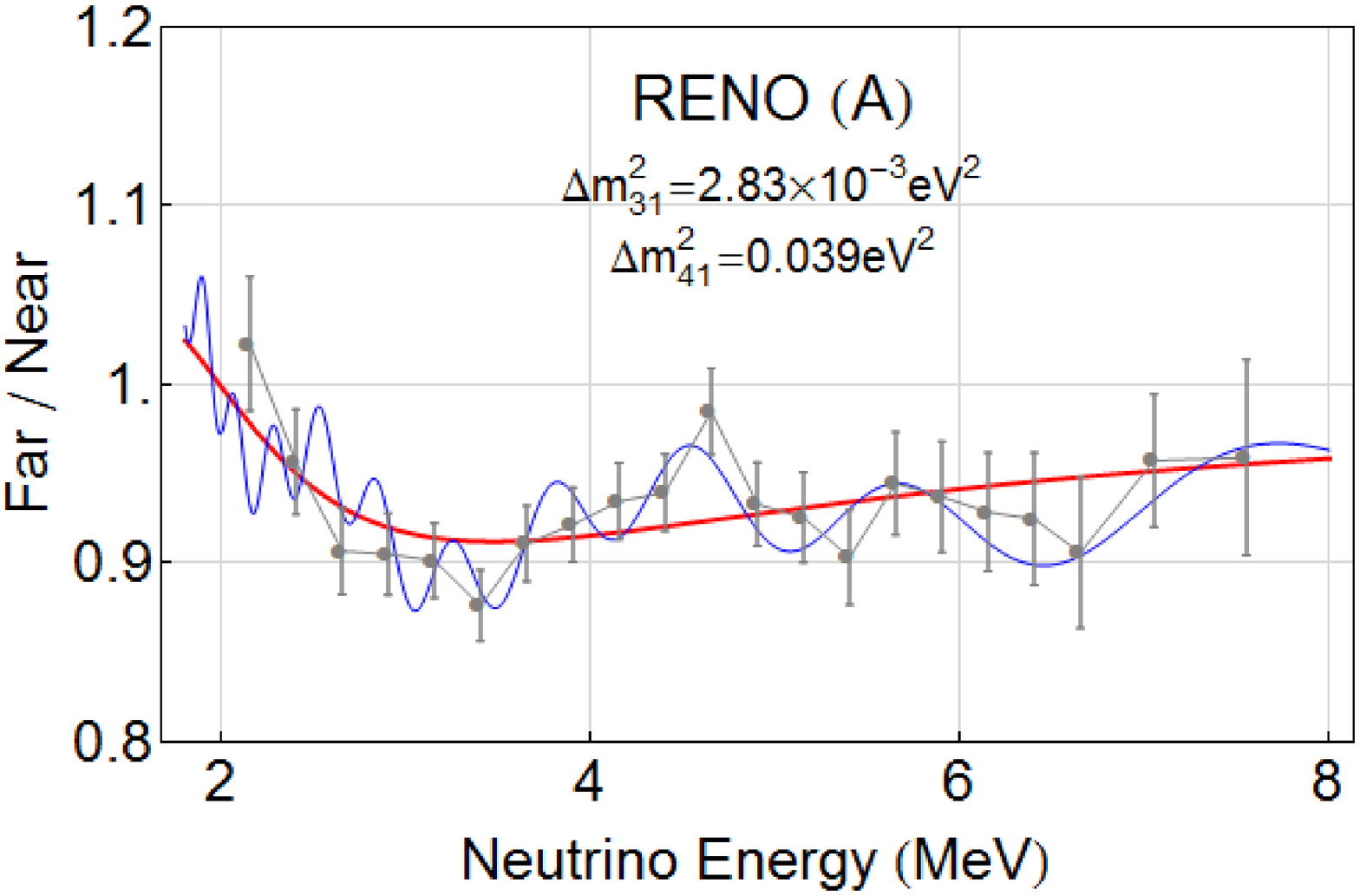}}
\resizebox{80mm}{!}{\includegraphics[width=0.75\textwidth]{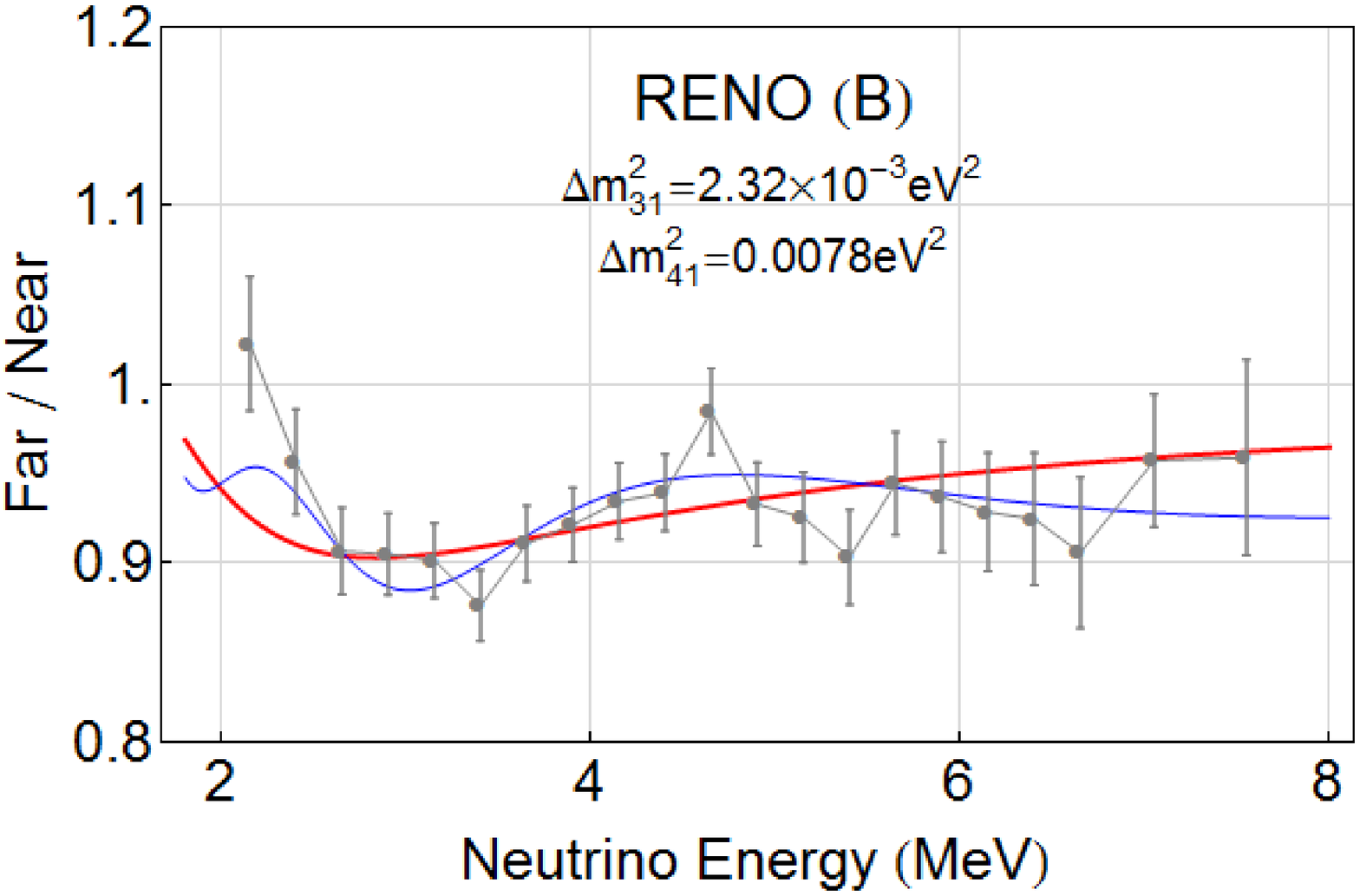}}
\caption{\label{fig:shape_analysis}
The data and error bars are the reproduction of updated RENO for an extended period until October, 2012. The red fits are drawn with $\sin^2 2\theta_{14}=0$ for $\Delta m_{31}^2=2.83\times10^{-3}\mathrm{eV}^2$ and for $\Delta m_{31}^2=2.32\times10^{-3}\mathrm{eV}^2$ obtained from the analysis in Fig. \ref{fig:3exclusion}. For each case, the blue fits are overlaid which are the superposition with $\Delta m_{41}^2=0.039 \mathrm{eV}^2$ oscillation of amplitude $\sin^2 2\theta_{14}=0.050$, and the superposition with $\Delta m_{41}^2=0.0078 \mathrm{eV}^2$ oscillation of amplitude $\sin^2 2\theta_{14}=0.054$, respectively, obtained from the analysis in Fig \ref{fig:exclusion}.}
\end{figure}

\begin{figure}
\resizebox{80mm}{!}{\includegraphics[width=0.75\textwidth]{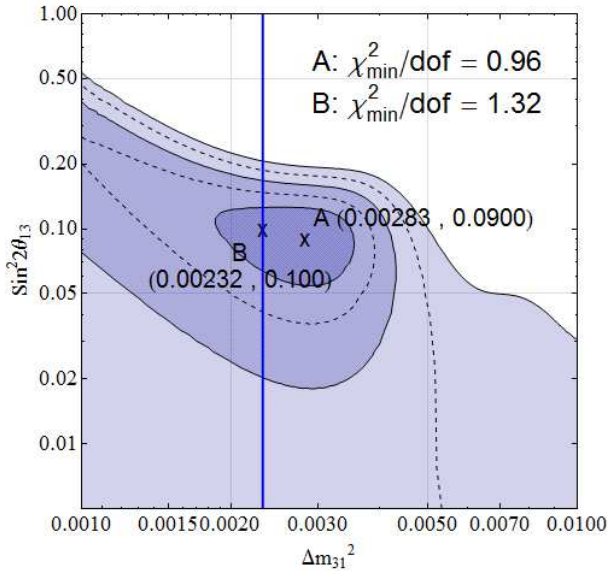}}
\caption{\label{fig:3exclusion}
Three-neutrino analysis of the spectral shape updated until October, 2012. The best fit in $(\Delta m^2_{31},~\sin^22\theta_{13})$ is (0.00283, 0.09) denoted by A. When $\Delta m^2_{31}=0.00232 \mathrm{eV}^2$ is fixed as taken by RENO, the best fit of $\sin^2\theta_{13}$ is 0.100.
The $1\sigma, ~2\sigma$ and $3\sigma$ exclusion curves are drawn by solid lines.}
\end{figure}

\begin{figure}
\resizebox{70mm}{!}{\includegraphics[width=0.75\textwidth]{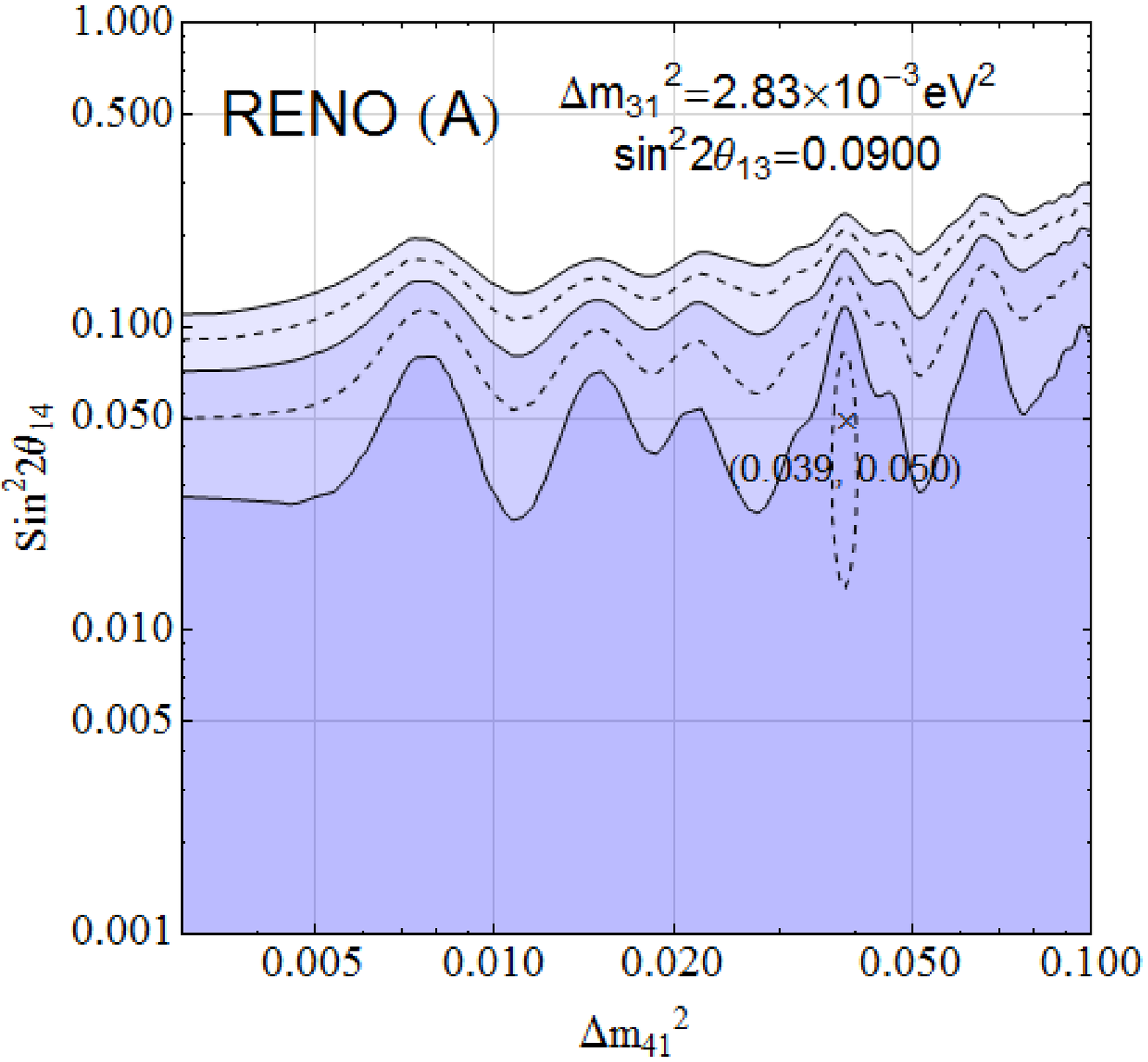}}
\resizebox{70mm}{!}{\includegraphics[width=0.75\textwidth]{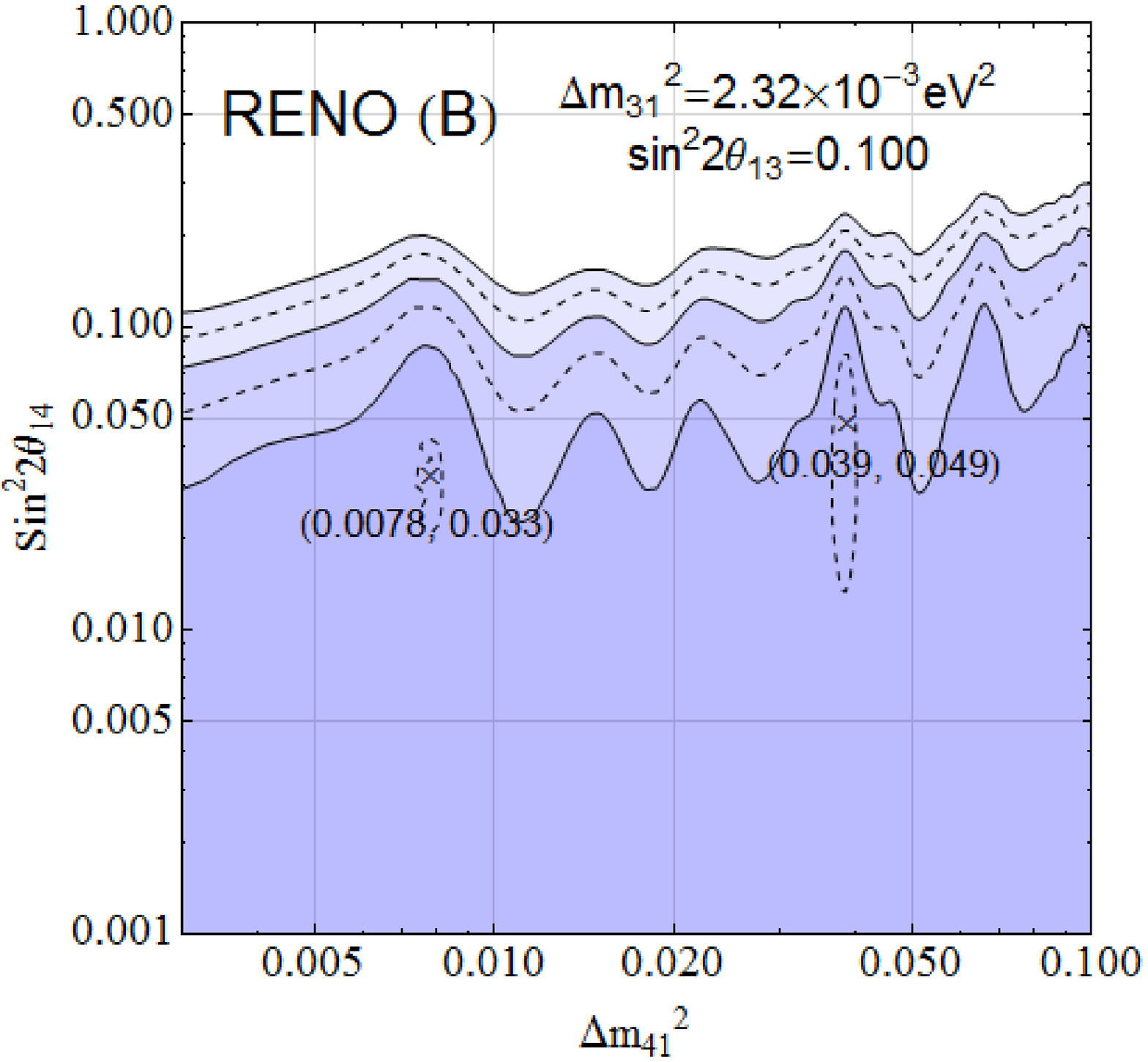}}
\caption{\label{fig:exclusion}
The $1\sigma, ~2\sigma$ and $3\sigma$ exclusion curves are drawn by solid lines. Apparently, a broad range of $\sin^2 2\theta_{14}$ apparently remains not excluded. Only $\Delta m^2_{41}>\Delta m^2_{31}$ is considered and $\sin^22\theta_{14}>0.2$ is excluded at $3\sigma$ CL. For (A) specified by $\Delta m_{31}^2=2.83\times 10^{-3}\mathrm{eV}^2$, the minimum $\chi^2_{\mathrm{min}}/\mathrm{dof}=0.48$ is at $(\Delta m^2_{41},~\sin^22\theta_{14})=(0.039\mathrm{eV}^2,~0.050)$, while, for (B) specified by $\Delta m_{31}^2=2.32\times 10^{-3}\mathrm{eV}^2$, two minima $\chi^2_{\mathrm{min}}/\mathrm{dof}=1.06$ and 0.85 are located at $(\Delta m^2_{41},~\sin^22\theta_{14})=(0.0078\mathrm{eV}^2,~0.033)$ and $(0.039\mathrm{eV}^2,~0.049)$, respectively.}
\end{figure}

\section{Four neutrino analysis of updated spectral shape in RENO}

One of RENO's results was the ratio of the observed to the expected number of antineutrinos in the far detector, $R=0.929 \pm 0.011$ (see Ref.\cite{seo:nutel}), where the observed is simply the number of events at FD. On the other hand, the expected number of events at FD can be obtained using several adjustments of the number of events at ND:
    \begin{eqnarray}
    R &\equiv&\frac{[\mathrm{Observed ~at ~FD}]}{[\mathrm{Expected ~at ~FD}]} \label{obs2exp}\\
     &\equiv& \frac{[\mathrm{No. ~of ~events ~at ~FD}]}{[\mathrm{No. ~of ~events ~at ~ND}]^*} ~, \label{far_to_near}
    \end{eqnarray}
where the number of events at each detector is normalized. The normalization of the neutrino fluxes at ND and FD requires an adjustment between the two individual detectors which includes corrections due to DAQ live time, detection efficiency, background rate, and the distance to each detector. The numbers of events at FD and ND in Eq.(\ref{far_to_near}) have already been normalized by these correction factors, and so we have $R_\mathrm{far}=0.929\pm0.017$ and $R_\mathrm{near}=0.990\pm0.025$ as shown in Fig.\ref{fig:theta13theta14}. The normalization guarantees $R=1$ at the center of the reactors. RENO removes the oscillation effect at ND when evaluating the expected number of events at FD by dividing the denominator of Eq. (\ref{far_to_near}) by 0.990 which is taken from $R_\mathrm{near}$. Now,
    \begin{eqnarray}
    R = \frac{[\mathrm{No. ~of ~events ~at ~FD}]}{[\mathrm{No. ~of ~events ~at ~ND}] ~/~0.990} ~, \label{jungsik}
    \end{eqnarray}
In rate-only analysis, the ratio of the observed to the expected number of events at FD in Eq.(\ref{obs2exp}) is just the survival at FD, since the denominator in Eq. (\ref{jungsik}) is eliminated. Thus, $R$ coincides with $R_\mathrm{far}$ in Fig. \ref{fig:theta13theta14}.

In spectral shape analysis, however, the denominator cannot be neglected, since the oscillation effect at ND differs depending on the neutrino energy. The data points in Fig.\ref{fig:shape_analysis} are obtained by the definition of the ratio $R$ given in Eq. (\ref{obs2exp}) and Eq. (\ref{far_to_near}) per 0.25MeV bin, as the energy varies from 1.8MeV to 12.8MeV. The data dots and error bars were updated by including additional data from March to October in 2012 officially announced at Neutrino Telescope 2013\cite{seo:nutel}. The ratio in Eq.(\ref{jungsik}) is compared with theoretical curves overlaid on the data points. The theoretical curves are described by
\begin{eqnarray}
\frac{\langle P(\mathrm{FD}) \rangle}{\langle P(\mathrm{ND})\rangle(0.990)^{-1}}, \label{far2near}
\end{eqnarray}
where $\langle P(\mathrm{L}) \rangle$ is given in Eq.(\ref{pee}). In Fig. \ref{fig:3exclusion}, the best fit of $(\Delta m_{31}^2, \sin^2 2\theta_{13})$ is presented when $\theta_{14}=0$. The point $\mathbf{A}~(0.00283\mathrm{eV}^2, 0.09)$ indicates the $\chi^2$ minimum where $\sin^22\theta_{13}$ and $\Delta m_{31}^2$ are parameters, while the point $\mathbf{B}~(0.00232\mathrm{eV}^2, 0.10)$ is the minimum where $\Delta m_{31}^2$ is 0.00232 which RENO and Daya Bay used for the fixed value. Hereafter, two cases depending on $\Delta m_{31}^2$ are discussed: One is for $\Delta m_{31}^2=0.00283\mathrm{eV}^2$ marked by \textbf{A} and the other is for  $\Delta m_{31}^2=0.00232\mathrm{eV}^2$ marked by \textbf{B}. According to the analysis performed with $\theta_{14}=0$, the red curves for the two values of $\Delta_{31}^2$ are overlaid on the spectral data in Fig. \ref{fig:shape_analysis}. Fig. \ref{fig:exclusion} shows interpretation of the spectral shape in terms of four-neutrino oscillation. For given values of $\Delta_{31}^2$ and $\sin^22\theta_{13}$, the 1, 2, 3$\sigma$ CL exclusion curves are obtained by convolution with the energy resolution of RENO detectors $(5.9/\sqrt{E}+1.1)\% $. Using the best fits $(\Delta m_{41}^2, \sin^2 2\theta_{14})=(0.039\mathrm{eV}^2, 0.05)$ for \textbf{(A)} and $(\Delta m_{41}^2, \sin^2 2\theta_{14})=(0.0078\mathrm{eV}^2, 0.033)$ for \textbf{(B)}, the blue curves are also added on the spectral data in Fig. \ref{fig:shape_analysis}. To see a distinct different aspect due to the magnitude of $\Delta m_{41}^2$, we choose $m_{41}^2=0.0078\mathrm{eV}^2$ for the best fit of \textbf{(B)}, avoiding $m_{41}^2=0.039\mathrm{eV}^2$ which is the same as the $m_{41}^2$ for \textbf{(A)}.

\begin{figure}
\resizebox{70mm}{!}{\includegraphics[width=0.75\textwidth]{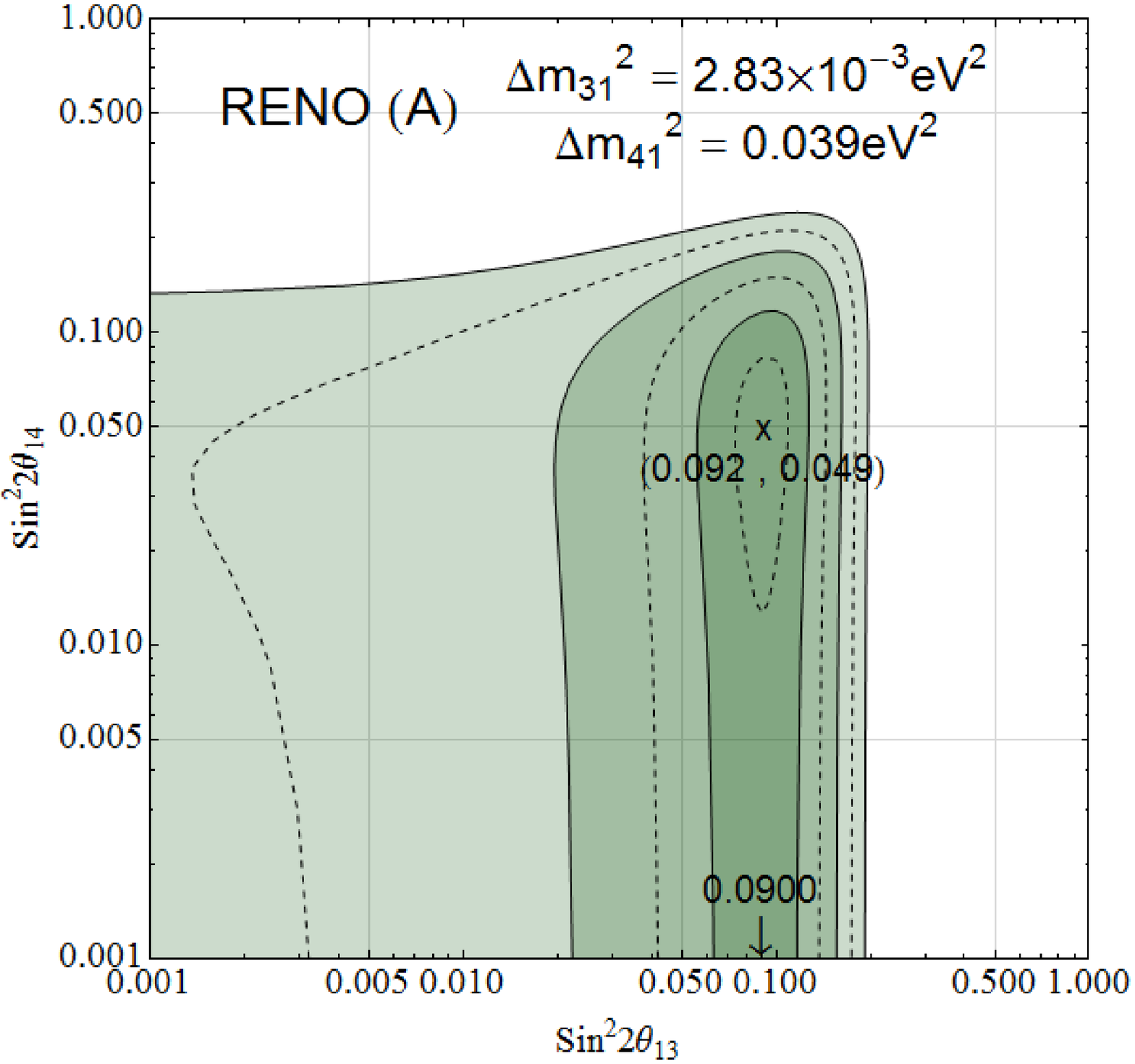}}
\resizebox{70mm}{!}{\includegraphics[width=0.75\textwidth]{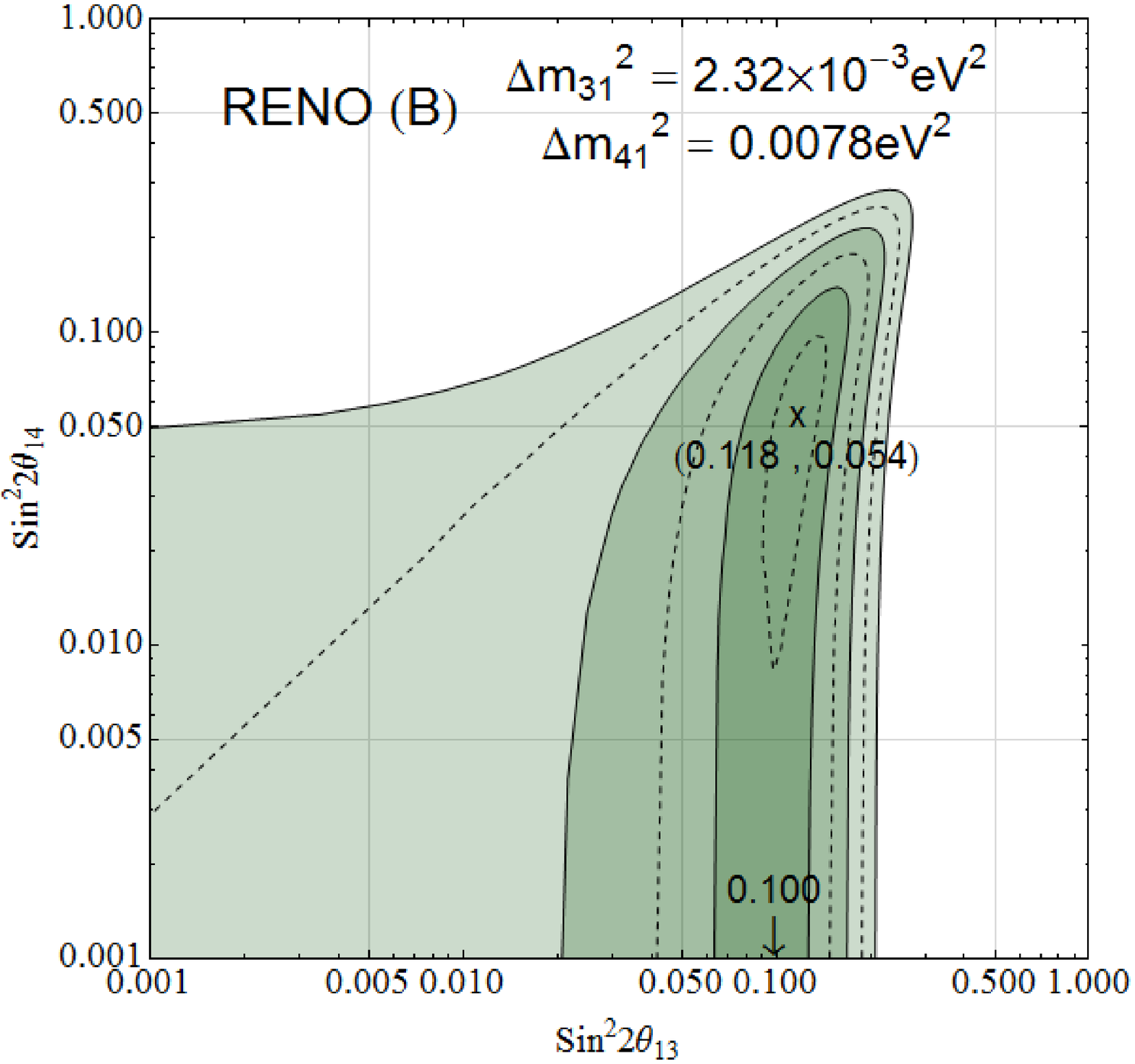}}
\caption{\label{fig:chi_sq}
The $1\sigma, ~2\sigma$ and $3\sigma$ fit of combination of the $\sin^2 2\theta_{13}$ and $\sin^2 2\theta_{14}$ for chosen values of $(\Delta m^2_{31}$ and $\Delta m^2_{41})$. For (A), the $\chi_{\mathrm{min}}^2/\mathrm{dof}$ of $(0.092,~0.049)$ is 0.51. For (B), the $\chi_{\mathrm{min}}^2/\mathrm{dof}$ of $(0.118,~0.049)$ is 0.96. In both cases, the best fit of $\sin^2 2\theta_{14}=0$ is included in 1$\sigma$ region.}
\end{figure}

The solid curves in Fig.\ref{fig:chi_sq} explain $1\sigma,~2\sigma$ and $3\sigma$ CL of $\Delta \chi^2$ in parameters $(\sin^22\theta_{13}, \sin^22\theta_{14})$, of which the best-fits are found at (0.092, 0.049) for case \textbf{(A)} of $\Delta m^2_{31}=0.00283\mathrm{eV}^2$ and at (0.118, 0.054) for case \textbf{(B)} of $\Delta m^2_{31}=0.00232\mathrm{eV}^2$. In case \textbf{(B)} where the value of $\Delta m^2_{31}$ is the same as the one that RENO and Daya Bay took for it, the best fit of $\sin^2 2\theta_{13}$ is 0.118 in company with non-zero $\sin^2 2\theta_{14}$. The best fit $\sin^2 2\theta_{13}=0.100$ with the restriction $\sin^2 2\theta_{14}=0$ is still within $1\sigma$ region of four-neutrino analysis. Also in case \textbf{(B)} which is specified by a rather large $\Delta m^2_{31}$ compared to the value taken by RENO and Daya Bay or the value suggested by global analyses, the best fit $\sin^2 2\theta_{13}=0.090$ of three-neutrino analysis is placed in the region of $1\sigma$ CL. This implies no preference between three-neutrino and four-neutrino schemes when the shape in Fig. \ref{fig:shape_analysis} is analyzed in this rough estimation.

\begin{figure}
\resizebox{70mm}{!}{\includegraphics[width=0.75\textwidth]{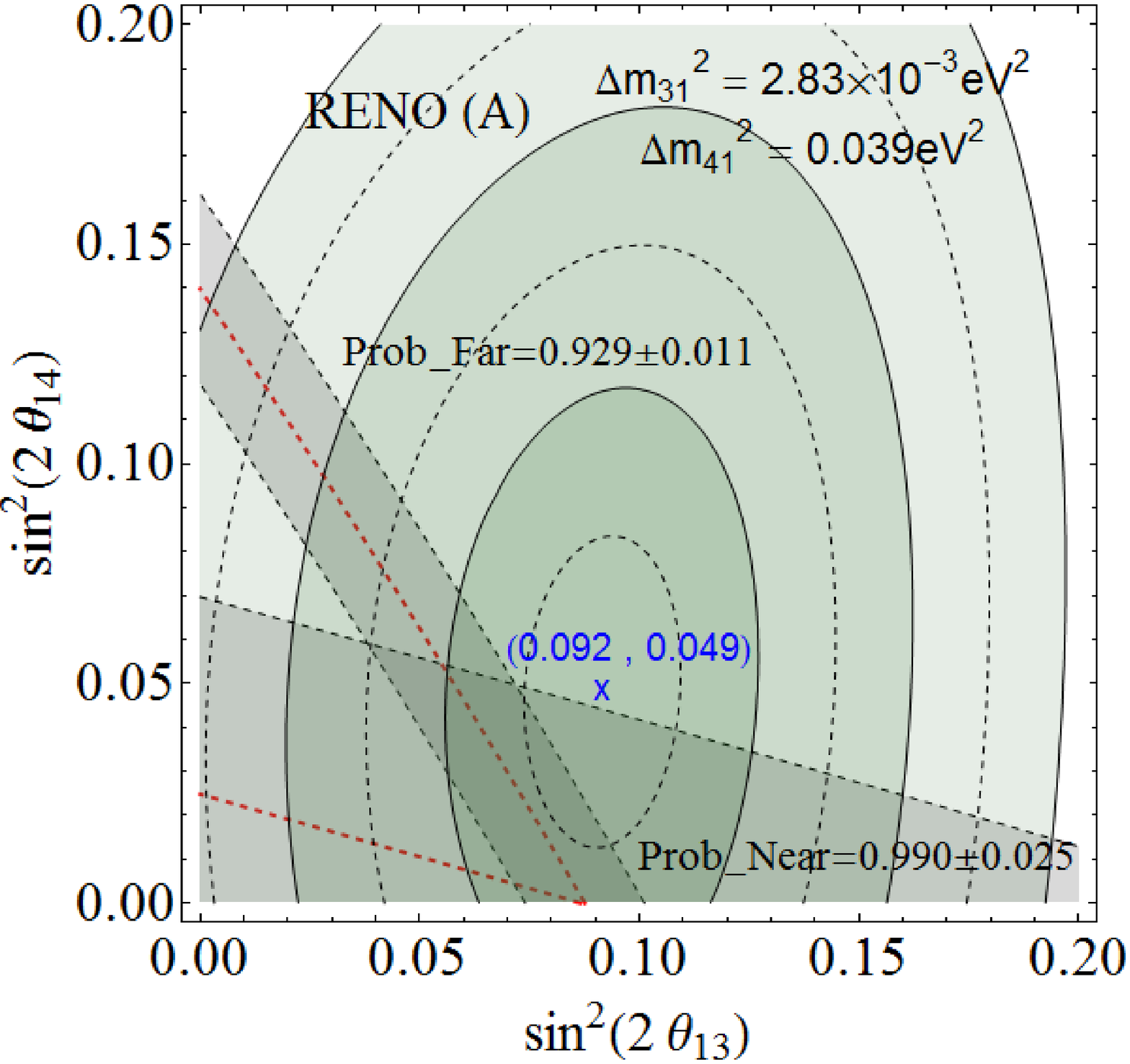}}
\resizebox{70mm}{!}{\includegraphics[width=0.75\textwidth]{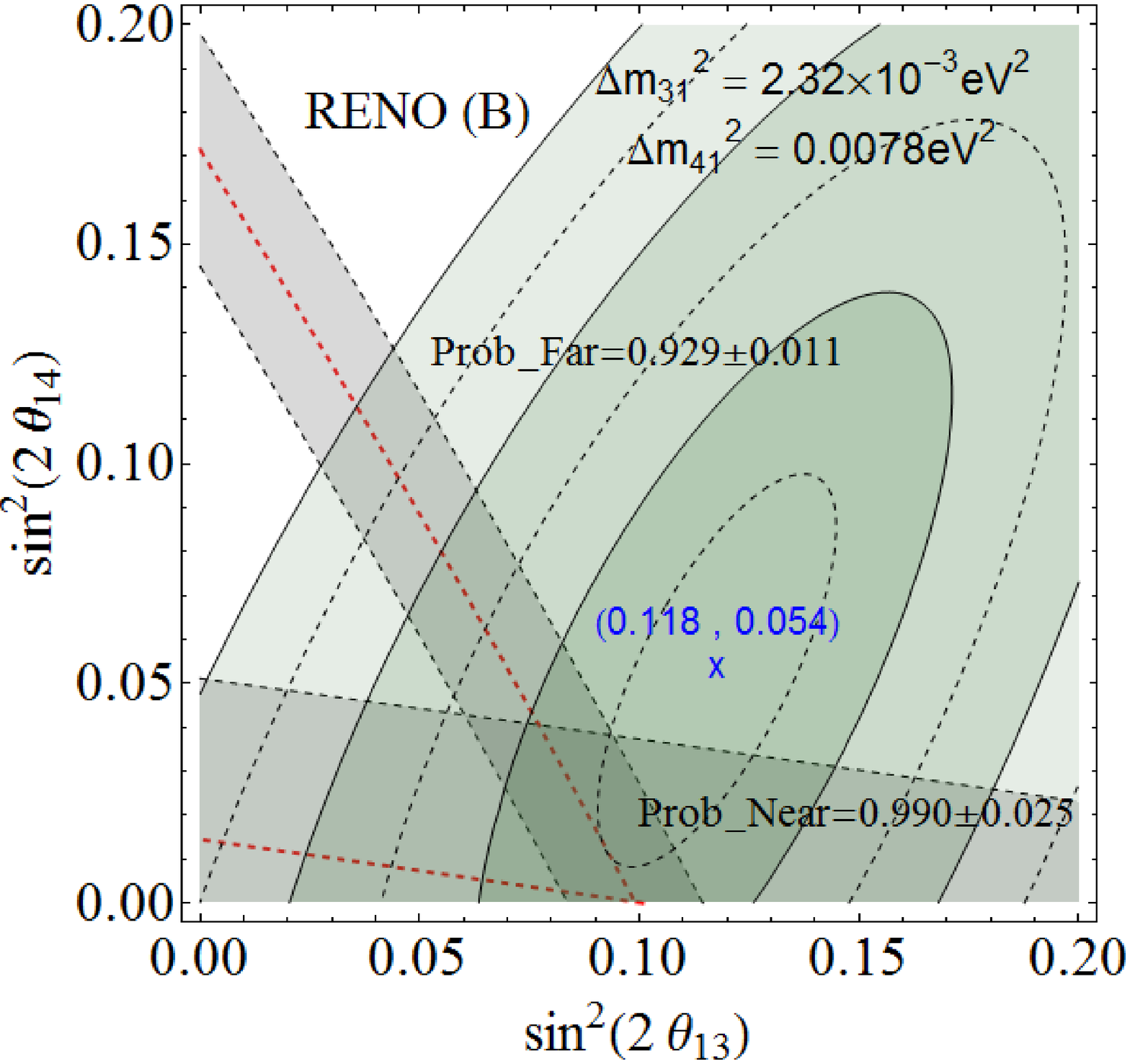}}
\caption{\label{fig:rate_shape}
Comparison of rate-only analysis and spectral shape analysis: In both (A) and (B), the regions allowed by rate-only analysis are overlapped in $1\sigma$ range of shape analysis, while they do not include the $\chi^2$-minimum best fits.}
\end{figure}

\section{Conclusion}

If a fourth type of neutrino has a mass not much larger than the other three masses, the results of reactor neutrino oscillations like RENO, Daya Bay, and Double Chooz can be affected by the fourth state. For detectors established for oscillations driven by $\Delta m_{31}^2=0.00232\mathrm{eV}^2$, clues about the fourth neutrino can be perceived only if the order of $\Delta m_{41}^2$ is not much larger than that of $\Delta m_{31}^2$. Therefore, this work examined the possibility of a kind of sterile neutrino in the range of mass-squared differences below $0.1\mathrm{eV}^2$, considering the two announced results of RENO and Daya Bay. Anomalies of reactor antineutrino oscillations have been considered for the range, $0.1\mathrm{eV^2}<\Delta m_{14}^2<1\mathrm{eV^2}$. Thus, it is worth analyzing the absolute flux at the near detector and the ratio of the far-to-near flux on a common basis\cite{reno:anomaly}.

RENO announced an update of rate-only analysis and the spectral shape of neutrino events[neutrino telescope], including an observed-to-expected ratio $R=0.929$ and an oscillation amplitude of $\sin^22\theta_{13}=0.100$. We compared the spectral shape with theoretical curves of the superpositions of $\Delta m_{41}^2$ oscillations and $\Delta m_{31}^2$ oscillations. In summary, $\sin^22\theta_{14}>0.2$ is excluded at $3\sigma$ CL. When $\Delta m_{31}^2=0.00232\mathrm{eV}^2$ is fixed, the best-fit in four-neutrino parameters is $(\Delta m_{41}^2, \sin^2 2\theta_{14})=(0.0078\mathrm{eV}^2, 0.054)$. When we search the fit of $\Delta m_{31}^2$ along with other parameters of four-neutrino analysis, the best value is obtained $\Delta m_{31}^2=0.00283\mathrm{eV}^2$ with $\sin^2 2\theta_{13}=0.090$ from the shape in three-neutrino analysis. When the parameters are extended to four-neutrino scheme, the best fit is $(\Delta m_{41}^2, \sin^2 2\theta_{14})=(0.039\mathrm{eV}^2, 0.049)$. As shown in Fig.\ref{fig:chi_sq}, the three-neutrino analysis of RENO $(\sin^22\theta_{13},~\sin^22\theta_{14})=(0.100,~0.0)$ is also included within $1\sigma$ CL in four-neutrino analysis. Thus, it is not yet known whether the superposition with $\Delta m_{41}^2$ oscillations is preferred to the single $\Delta m_{31}^2$ oscillations at RENO  detectors. Fig. \ref{fig:rate_shape} shows that the rate-only analysis and the spectral shape analysis are in good agreement within their $1\sigma$ CL range.

\begin{acknowledgments}
This research was supported by a Chung-Ang University Research Scholarship grant in 2013. KS was supported by the National Research Foundation of Korea(NRF) grants funded by the Korea Government of the Ministry of Education, Science and Technology(MEST) (2011-0014686).
\end{acknowledgments}

\appendix

\end{document}